\documentclass[superscriptaddress,twocolumn]{revtex4-1}
\usepackage{graphicx}
\usepackage{amsmath}
\usepackage{longtable}
\usepackage{amssymb}
\usepackage{latexsym}
\usepackage{times}
\usepackage{graphics}
\usepackage{graphicx}
\usepackage{epsfig}
\usepackage{wrapfig}
\usepackage[T1]{fontenc}
\usepackage{float}
\usepackage{enumerate}
\usepackage{setspace,ifthen}
\usepackage{amsthm} 
\usepackage{amssymb}	
\usepackage[usenames,dvipsnames]{xcolor}
\usepackage{mathrsfs,amsmath}
\usepackage{mathrsfs}
\usepackage{mathrsfs}
\usepackage{rotating}
\usepackage{epstopdf}
\usepackage{appendix}
\usepackage{blkarray}
\usepackage{multirow}

\usepackage{array}

\usepackage{rotate}

\newcommand{\abs}[1]{\left| #1 \right|} 

\let\baraccent=\= 
\renewcommand{\=}[1]{\stackrel{#1}{=}} 

\theoremstyle{definition}

\theoremstyle{remark}

\newcommand{\sigz}{\hat{\sigma}_z}
\newcommand{\sigp}{\hat{\sigma}_+}
\newcommand{\sigm}{\hat{\sigma}_-}

\newcommand{\E}[2][1]{\left\langle #2\right\rangle}

\newcommand{\Htot}{ {H}}
\newcommand{\Hq}{ {H}_0}

\newcommand{\sigvec}{\boldsymbol{\sigma}}

\newcommand{\hvec}{\boldsymbol{h}}

\renewcommand{\dh}[2][1]{\delta h_{#2}}

\newcommand{\sigx}{\hat{\sigma}_x}
\newcommand{\sigy}{\hat{\sigma}_y}

\newcommand{\HphiN}{H_{\dot{\phi}_N}}

\newcommand{\fnoise}{{\delta\omega_\text{LO}}}
\newcommand{\SOS}{S^{(1)}}

\newcommand{\wLO}{\omega_\text{LO}}
\newcommand{\Qsplitting}{\omega_0}

\begin{document}



\title{The role of master clock stability in scalable quantum information processing}
\author{Harrison Ball}
\affiliation{ARC Centre for Engineered Quantum Systems, School of Physics, The University of Sydney, NSW 2006 Australia}
\author{William D. Oliver}
\affiliation{Department of Physics, Massachusetts Institute of Technology, Cambridge, MA 02139 USA}
\affiliation{MIT Lincoln Laboratory, Lexington, MA 02420 USA}
\author{Michael J. Biercuk}
\affiliation{ARC Centre for Engineered Quantum Systems, School of Physics, The University of Sydney, NSW 2006 Australia}
\email[To whom correspondence should be addressed: ]{michael.biercuk@sydney.edu.au}

\date{\today}
\begin{abstract}
Experimentalists seeking to improve the coherent lifetimes of quantum bits have generally focused on mitigating decoherence mechanisms through, for example, improvements to qubit designs and materials, and system isolation from environmental perturbations. In the case of the phase degree of freedom in a quantum superposition, however, the coherence that must be preserved is not solely internal to the qubit, but rather necessarily  includes that of the qubit relative to the ``master clock'' (e.g. a local oscillator) that governs its control system. In this manuscript we articulate the impact of instabilities in the master clock on qubit phase coherence, and provide tools to calculate the contributions to qubit error arising from these processes.  We first connect standard oscillator phase-noise metrics to their corresponding qubit dephasing spectral densities.  We then use representative lab-grade and performance-grade oscillator specifications to calculate operational fidelity bounds on trapped-ion and superconducting qubits with relatively slow and fast operation times. We discuss the relevance of these bounds for quantum error correction in contemporary experiments and future large-scale quantum information systems, and discuss potential means to improve master clock stability.

\end{abstract}



\maketitle



\section{Introduction}\label{Sec:Intro}
A fundamental challenge to the broader application of quantum information science is the management of error in fragile quantum hardware~\cite{NC, oskin2002practical}. The need for higher-fidelity performance motivates research at all architectural levels~\cite{JonesPRX2012}, from theoretical studies of fault tolerance and analyses of quantum error correction (QEC) implementations down to experimental improvements in the operational fidelity of elemental devices.  A familiar aspect of this challenge is decoherence, a process by which even idle qubits undergoing free evolution (i.e., the identity operator) will gradually lose their stored quantum information, rendering them useless in subsequent computation.

A prevalent component of decoherence is dephasing, a randomization of the relative phase between the basis states that form a coherent superposition state.  Dephasing in qubit systems is commonly attributed to environmental fluctuations of a qubit bias or control parameter, e.g., an external magnetic field, which modulates the qubit-state energy splitting and hence its dynamic phase evolution.  In turn, experimentalists have primarily focused their efforts on reducing both the level of environmental noise and the qubit sensitivity to that which remains (c.f.~\cite{Houck2009, RevModPhys.79.1217, Langer2005}).  As a result, many qubit technologies~\cite{Ladd2010} have witnessed remarkable performance improvements, with dephasing times approaching milliseconds or even seconds (depending on the qubit modality)~\cite{Langer2005, OlmschenkPRA2007, SoarePRA2014, Chow2012, SellarsPRA, SellarsLuminescence, Morello}, and operational fidelities reaching $\sim99.999\%$~\cite{BrownPRA2011, MartinisOptimized, Lucas2014}.  Achieving targets such as these - once thought all but impossible -  has led many to recognize that scalable, error-corrected quantum computation may become a reality.  In essence, by improving qubit coherence, error rates may be reduced below fault-tolerant thresholds, and in principle QEC may be employed to suppress errors at arbitrary system scales.

This prescription, however, belies issues that only arise as qubit coherence and operational fidelity improve. A fundamental example, one widely studied in the context of precision metrology, is that the control and measurement of highly coherent systems must be compared to a suitably coherent reference~\cite{audoin2001}. Qubit coherence, likewise, is inferred relative to a reference, generally in the form of a local oscillator (LO) used to control and interrogate the system, for instance by inducing Ramsey or Rabi oscillations.  Here, the evolution of the qubit's phase degree of freedom in a quantum superposition is effectively being compared against the accumulated ticks of a clock, defined by the LO~\cite{audoin2001}.  While the importance of phase coherence is known to any experimentalist who has failed to lock an oscillator to a stable reference, in the past, qubit coherence times and operational fidelities have generally been limited by environmental noise.  At today's performance levels, however, noise in the master clock is beginning to emerge as a contributor to the overall error rate.  And, as qubits continue to improve, mitigating master clock instabilities will become material to high fidelity operation.

This manuscript addresses the topic of master clock instability and its impact on qubit coherence and operational fidelities for quantum information applications. We begin by accounting for the master clock in a Hamiltonian treatment of qubit dephasing.  From this starting point, we engage in an analysis of the relevance of master-clock instabilities to qubit coherence and dephasing-induced error rates.  We identify master-clock phase fluctuations as an emerging source of error in today's qubit systems and one that will certainly become more prominent as qubit error rates continue to improve.  Our presentation serves to unify concepts familiar to quantum information~\cite{NC}, quantum control~\cite{BiercukQIC2009, BiercukJPB2011, SoarePRA2014}, engineering~\cite{rutman1978, GreenNJP2013}, and precision frequency metrology~\cite{nist1990} in order to allow the translation of LO phase noise specifications -- as presented in experimental data sheets -- into gate error probabilities for a variety of canonical single-qubit operations.  We consider gates applied to superconducting and trapped-ion qubits, prominent qubit modalities that represent systems with relatively fast (10 ns) and slow (100 $\mu$s) gate times yielding sensitivity to different frequency regions of the oscillator noise spectrum. Calculations presented in this analysis highlight the fact that qubit dephasing errors arising from the phase fluctuations in commercial precision LOs, while not explicitly limiting experiments today, will likely become a significant consideration in the context of large-scale quantum information.  They also reveal that many lab-grade oscillators are beginning to limit achievable fidelities in contemporary systems.  We address the relevance of different frequency regimes of LO phase noise, and highlight ``far-from-carrier'' phase noise as contributing to important upper-bounds on qubit operational fidelities.   The materials we present both demonstrate a path to mesoscale quantum information systems using existing master clocks and provide new motivation for investment in LO hardware and precision frequency metrology research in order to underpin large-scale quantum information processing.  We augment this review with detailed supplementary material, aggregating a comprehensive theoretical foundation to understand clock-induced errors in quantum systems.

\section{Qubit dephasing induced by the master clock}\label{Sec:Theory}
In a semiclassical picture, qubit phase coherence corresponds to maintaining the dynamical phase of the qubit relative to a reference (master) clock for the system.
Implicit in the Bloch sphere representation of a qubit state is, in general, a transformation to a frame co-rotating with the nominal qubit Larmor frequency, $\omega_{0}$.  %
%
In a dephasing process, the relative phase difference between the qubit state and its reference frame evolves stochastically in time, introducing a degree of randomness to the qubit state.  Typical formulations assign this stocastic evolution to instability in qubit frequency, induced, for instance, by fluctuations in external fields. Over time, these fluctuations cause the qubit to randomly advance or retreat relative to the co-rotating frame.  However, what about the stability of the rotating frame being used as a phase reference?  The simple observation that the qubit phase is defined relative to the rotating frame indicates that an experimentalist must consider not only the stability of the qubit, but also the stability of the rotating frame itself, realized through the use of a local oscillator in the experiment.  Accordingly, the master clock has a role that is more fundamental than the synchronization of scheduled operations; the master clock determines, in part, the coherence of the underlying qubits.

The impact of generic qubit dephasing may be formally captured through a Hamiltonian formulation, written as the sum of an ideal control component and a randomly fluctuating noise component in the three-dimensional Pauli basis~\cite{Hamiltonian}.
\begin{align}\label{Eq:HamiltonianIncludingLO&environmentalDephasing}
    \Htot(t) &=\hvec(t)\boldsymbol{\sigma}+\dh{z}(t)\sigz,\\
    \dh{z}(t) &= \dh{z}^{(env)}(t)+\dh{z}^\text{(LO)}(t)
\end{align}
where we have restricted attention to longitudinal dephasing noise by setting $\dh{x,y}(t) = 0$.  Terms from environmental dephasing $\dh{z}^{(env)}(t)$ and LO-induced dephasing, $\dh{z}^\text{(LO)}(t)$, appear on an equal footing in this formulation
(for full derivation, see \emph{Methods} and \emph{Supplementary Information}).
Once a relative phase relationship between the qubit and LO is established via an initial interaction
(control pulse), the presence of these two noise terms accounts for qubit dephasing at all times $t$.

Within this framework, we can build a connection between the LO-induced dephasing term and the noise properties of local oscillators.  A local oscillator with amplitude $A(t)$ and carrier frequency $\wLO$
possesses a phase that is easily partitioned into a ``control'' component (desired, deterministic phase evolution), $\phi_C(t)$,  and a ``noise'' component (unwanted, stochastic phase evolution), $\phi_N(t)$, in the expression for the oscillator signal: $A(t)\cos\Big(\wLO t +[\phi_{C}(t)+\phi_{N}(t)]\Big)$.
Setting the LO frequency to be on resonance with the qubit, $\wLO=\omega_{0}$, the stability of the rotating frame is determined by the noisy phase evolution of the LO, $\phi_N(t)$, as it produces a time-dependent frequency detuning of the LO relative to the qubit through its time derivative, $\fnoise(t)\equiv\dot{\phi}_N(t)$.  With this qualitiative understanding in hand, we are able to associate the Hamiltonian term $\dh{z}^\text{(LO)}(t)=-\frac{1}{2}\dot{\phi}_N(t)$, explicitly linking LO phase fluctuations to a dephasing Hamiltonian
with an impact that is indistiguishable from other environmental dephasing sources.

\begin{figure}[t]
\centering
\includegraphics[width=9cm]{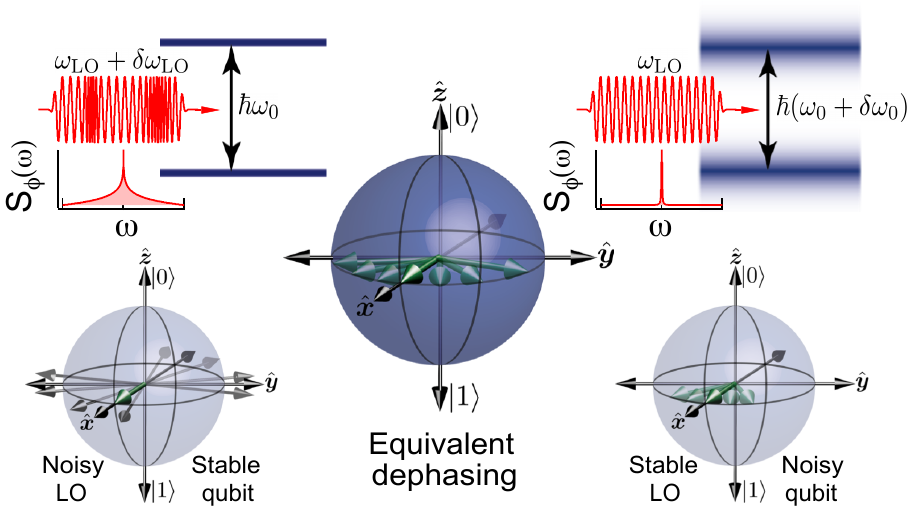}
\caption{Schematic representation of the correspondence between phase fluctuations and environmental dephasing.  A stable qubit interacting with a LO experiencing frequency instability (captured by the phase instability $S_{\phi}(\omega)$ - the power spectral density of phase fluctuations), induces random phase accumulation in a freely evolving qubit, represented in the Bloch sphere picture.  Phase instability is frequently represented in the fourier domain as a broadened lineshape of the LO's carrier.  The same phenomenology arises for a LO outputting a perfectly stable sinusoidal signal in the presence of environmental Hamiltonian terms that produce fluctuations in the frequency of the qubit transition.}
\label{Fig:BlochSphereEquivalenceConcept}
\end{figure}

In Fig.~\ref{Fig:BlochSphereEquivalenceConcept} the equivalence of the two dephasing terms described above is represented on the Bloch sphere in the frame co-rotating with the resonant driving field.
The term $\dh{z}^{(env)}(t)$ induces rotations of the qubit Bloch vector about the $\hat{z}$-axis of the Bloch sphere, relative to a fixed coordinate frame set by an ideal LO.  Conversely, LO instability captured by $\dh{z}^\text{(LO)}(t)$ induces rotations of the coordinate frame relative to a fixed Bloch vector representing an ideal qubit.  From the perspective of the qubit, both terms are sources of dephasing. Hereafter, without loss of generality, we will ignore environmental dephasing by setting $\dh{z}^{(env)}(t) = 0$ in order to restrict attention to LO-induced dephasing.

\section{Calculating dephasing due to phase noise in the local oscillator}
We employ the filter transfer function formalism to incorporate
noise power spectral densities for the Hamiltonian noise terms into calculations of operational fidelity and qubit coherence~\cite{GreenNJP2013}.
This approach treats the problem of calculating the
impact of noise on a quantum system in terms of the overlap of the noise and an effective frequency-domain filter describing the action of the control~\cite{KurizkiPRL2001, KurizkiPRL2004}.
It adopts concepts that are well known in the engineering literature~\cite{Stengel} and has recently been developed and applied to the control of qubits~\cite{BiercukNature2009, BiercukPRA2009, Bylander2011, SoareNatPhys2014}.

In essence, the filter transfer function shapes the spectrum of the underlying noise as it couples to the qubit, passing certain bands and rejecting others.  We see this quantitatively for a system Hamiltonian in the form of Eq. \ref{Eq:HamiltonianIncludingLO&environmentalDephasing} by writing the fidelity of an arbitrary qubit operation 
in terms of the overlap integral of $\SOS_{z}(\omega)$, the unilateral (one-sided) power spectral density (PSD)~\cite{PSDs} of the dephasing field $\dh{z}(t)$, and the filter transfer function $G_{z,l}(\omega)$, 
which captures the action of the control Hamiltonian $\hvec(t)\sigvec$~\cite{GreenPRL2012, GreenNJP2013}.  We express the average fidelity

\begin{align}\label{Eq:DecoherenceChi}
\mathcal{F}_{av}(\tau)&\approx\frac{1}{2}\big\{1+\exp[-\chi(\tau)]\big\},\\
\label{Eq: DecoherenceChi}\chi(\tau)&=\left(\frac{1}{\pi}\right)\int_{0}^{\infty}\frac{d\omega}{\omega^2} \SOS_{z}(\omega)\sum_{l\in\; x,y,z} G_{z,l}(\omega).
\end{align}

\noindent The fact that the error integral, $\chi(\tau)$, is expressed as a product of the noise power spectral density and filter transfer functions demonstrates how the frequency-domain shaping of the noise determines the ultimate contribution of the noise to operational infidelity.  The transfer functions for the control, $G_{z,l}(\omega)$, may be calculated analytically and have contributions along all Cartesian directions (indexed by $l$), as dephasing noise present during a non-commuting control operation (e.g. $\propto\hat{\sigma}_{x}$) induces both dephasing and amplitude-damping~\cite{GreenPRL2012, GreenNJP2013}.
Alternatively, the $G_{z,l}(\omega)$ corresponding to a Ramsey pulse sequence would determine dephasing during free evolution.

Our objective is to link noisy fluctuations in the LO phase to the quantity $\SOS_{z}(\omega)$ relevant to the Hamiltonian noise terms.  Since we are restricting attention to the case $\dh{z}(t) = \dh{z}^\text{(LO)}(t)$ (omitting environmental noise for convenience), the PSD $\SOS_{z}(\omega)$ will solely describe fluctuations in the LO phase.  For this we reference a large base of research from the field of precision metrology, where the characterization of time and frequency signals is a key objective~\cite{audoin2001}.  Conveniently, in this discipline the stability of a signal at a notionally fixed frequency is quantified by characterizing temporal fluctuations in the phase of that signal, represented in the Fourier domain. Similar analyses are found in the context of communications engineering and stochastic signal processing.

The quantity of metrological significance~\cite{nist1990} in LO characterization is
$\SOS_{\phi_{N}}(\omega)\equiv \lim_{T\rightarrow\infty}\frac{2}{T}\E{{\abs{\Phi_{T}(\omega)}^2}}$,
the unilateral power spectral density of the LO phase fluctuations over a measurement time $T$.
In this expression,
$\Phi_{T}(\omega)=\int_{-T/2}^{T/2}\phi_{N}(t)e^{-i\omega t}dt$
is the complex amplitude of the harmonic Fourier component at frequency $\omega$ for the time-gated signal $\phi_N(t)$, defined for times $|t| < T/2$ and zero otherwise.  The quantity $\abs{\Phi_{T}(\omega)}^2$ is the energy density of this harmonic with units of energy per Hertz; angle brackets indicate an expectation value; and stationarity has been assumed to take the limit $T\rightarrow\infty$.
The phase fluctuations generate sidebands on the carrier -- the unadulterated LO signal --
at frequencies $\omega_\text{LO}\pm\omega$, effectively broadening the observed linewidth of the LO in frequency space (see Fig.~\ref{Fig:BlochSphereEquivalenceConcept}).
The Fourier harmonic $\omega$ of the phase instability is therefore measured as an offset from the carrier, and the power of these sidebands is measured 
over a 1 Hz bandwidth.
\noindent

\begin{figure}[t]
\centering
\includegraphics[width=0.9\columnwidth]{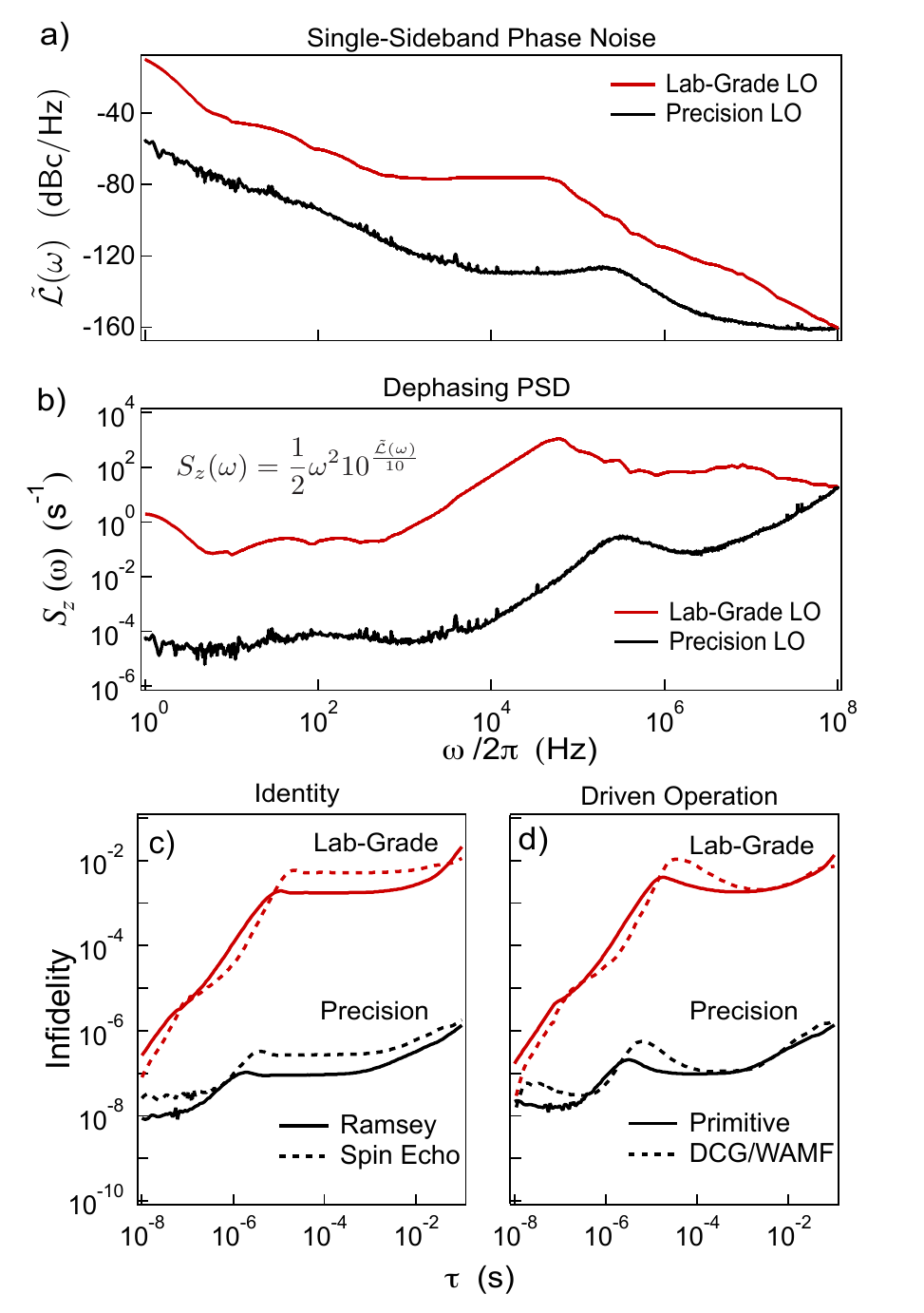}
\caption{a) Local oscillator phase noise expressed as $\tilde{\mathcal{L}}(\omega)$ for two grades of synthesizer, b) converted to $\SOS_{z}(\omega)$ via Eq.~\ref{Eq:ScaledFrequencyVsPhasePSD}.   Data extracted from specification sheets and manufacturer measurements.  For the ``lab-grade" synthesizer, phase noise at frequencies $\omega/2\pi<10$ Hz and $\omega/2\pi>10$ MHz has been extrapolated following the trending behaviour of manufacturer-supplied data for $\tilde{\mathcal{L}}(\omega)$. c-d) Calculated infidelity for different classes of quantum logic operations (see main text).  We have numerically confirmed that, for evolution times $\tau<100$~ms, the presence of a sharp low-frequency cutoff below $\omega/2\pi<1$ Hz contributes a negligible correction to our calculated results based on extrapolating $\tilde{\mathcal{L}}(\omega)$ as described above.}
\label{Fig:InfidelityCurves}
\end{figure}

The time-derivative of the phase fluctuations is equivalent to a time-dependent detuning of the LO from resonance, as described above, and we may therefore relate their noise spectra to the dephasing power spectrum used in filter-function calculations as $\frac{1}{4} \omega^2 \SOS_{\phi_{N}}(\omega)= \SOS_{z}(\omega)$.   The factor 1/4 arises from omitting the counter-rotating term in the rotating wave approximation. While $\SOS_{\phi_{N}}(\omega)$ is used in the metrological community~\cite{nist1990}, most LO manufacturers use a metric for the single-sideband phase noise, $\tilde{\mathcal{L}}(\omega)=10\log_{10}[\frac{1}{2}\SOS_{\phi_N}(\omega)]$, expressed in logarithmic units of dBc/Hz.  The ultimate relationship therefore allows tabulated single-sideband phase-noise specifications to be converted to unilateral dephasing power spectral densities as
\begin{align}\label{Eq:ScaledFrequencyVsPhasePSD}
    \SOS_{z}(\omega)=\frac{1}{2} \omega^2 10^{\frac{\tilde{\mathcal{L}}(\omega)}{10}}.
\end{align}
This correspondence enables a connection between the metrics provided by oscillator manufacturers
and those used for both precision oscillator characterization and for qubit dephasing calcuations~\cite{nist1990}, and it is represented graphically in the difference between curves in Figs.~\ref{Fig:InfidelityCurves}a-b.
A full accounting of all relevant factors and the formal derivation of these quantities is presented in the \emph{Supplementary Information} in a single consistent notation.

\section{Clock errors induced by state-of-the-art Local Oscillators}
With Eq~\ref{Eq:ScaledFrequencyVsPhasePSD} in hand, it is possible to use LO phase-noise specifications found on instrumentation data sheets for calculations of qubit coherence and operational error rates.
We calculate the operational infidelity $1-\mathcal{F}_{av}(\tau)$ using published $\tilde{\mathcal{L}}(\omega)$ for representative synthesizers
and analytic filter transfer functions as inputs for Eq. \ref{Eq: DecoherenceChi}.
We consider two standard operations:
the identity operation, $\hat{\mathbb{I}}$, and the $\hat{X}$ gate, defined as a driven rotation through angle $\pi$ about the $\hat{x}$-axis of the Bloch sphere.
%
These choices are representative of the
single-qubit operations comprising a universal gate set for gate-based quantum computation.
The former allows insight into the coherent lifetime of the qubit under free evolution ($T_{2}$), while the latter informs how gate fidelity is reduced during driven evolution by the presence of LO-induced dephasing noise.
For each of these ``primitive'' operations, the corresponding filter transfer function $G_{z,l}(\omega)$ is calculated following the techniques presented in Refs.~\onlinecite{Martinis2003, UhrigNJP2008, BiercukJPB2011,Fei_2012, GreenPRL2012, GreenNJP2013, SoareNatPhys2014}.

In addition to the primitive forms of these operations, we also employ specific dynamic error suppression (DES) strategies~\cite{Viola1998,Viola1999,Zanardi1999,Vitali1999,Viola2003,Byrd2003,KurizkiPRL2004,Khodjasteh2005,Yao2007,Uhrig2007,Gordon2008,Khodjasteh2009dcg,Khodjasteh2009,Khodjasteh2010,Liu2010,Biercuk_Filter} designed to reduce errors due to dephasing noise.
Such controls generally constitute time-dependent modulation of the system dynamics with the aim of coherently averaging out slow fluctuations.
While these protocols have typically been associated with the mitigation of environmental decoherence, as the reader might expect based on the discussion in Section II, a growing body of literature has shown that errors induced by imperfect control -- including LO phase noise -- can also be mitigated using these same techniques~\cite{BiercukNature2009, SoarePRA2014, SoareNatPhys2014}.
Here, for the identity operation (free evolution), we use the simple example of a spin echo, while for the driven
$\hat{X}$ gate, we employ a dynamically corrected version of this gate~\cite{Khodjasteh2009,Khodjasteh2009dcg, Khodjasteh2010}, also described as a Walsh amplitude modulated filter~\cite{SoareNatPhys2014, Ball2014}.
A description and analytic expressions for the filter functions used here may be found in the \emph{Supplementary Information}.

For this analysis, we have selected two distinct grades of LO to demonstrate the significance of LO choice in quantum control experiments.
Figure ~\ref{Fig:InfidelityCurves}a shows the single-sideband phase noise at 10 GHz for both a ``lab grade'' synthesizer (Vaunix LMS-123) and a ``precision'' synthesizer (Keysight/Agilent 8267D OPT-UNY).
Across much of the band, as expected, the phase noise for the precision synthesizer is lower -- by approximately 40 dB -- than the lab-grade unit.
For the LOs studied here $\tilde{\mathcal{L}}(\omega)$ declines rapidly with offset frequency from the carrier, exhibiting various power-law-dependences over the entire frequency range.

Generically, these dependences are captured analytically by the form $\tilde{\mathcal{L}}(\omega)\propto\omega^{p}$ ({\emph{i.e.},} $S_{z}(\omega)\propto\omega^{p+2}$).
Moving away from the carrier, dominant processes common in LOs include random-walk frequency noise ($p=-4)$, flicker frequency noise ($p=-3$), white frequency noise ($p=-2$), flicker phase noise ($p=-1$), and white phase noise ($p=0$).
Note that as expected, the dephasing noise mechanisms carry an exponent $p+2$ with respect to the corresponding phase noise mechanisms with exponent $p$ (Fig.~\ref{Fig:InfidelityCurves}b).
Details of the origins of the underlying processes may be found in~\cite{rutman1978}.


Calculations of infidelity for the four operations outlined above are presented in Fig.~\ref{Fig:InfidelityCurves}c-d and summarized in Table~\ref{Table:T1}.
For the identity operator, we find an approximate improvement of $10^{4}$ in residual error rate due to the use of a high-precision frequency source, and the infidelity remains below $10^{-6}$ out to 100 ms evolution time.
In contrast the lab-grade synthesizer induces an error exceeding $\sim0.1\%$ beyond a few microseconds of evolution time.
%
%
The plateau-like behavior in infidelity with increasing evolution time is due to the interplay of the dephasing power spectrum and filter-function~\cite{LongStorage}, and is similar to phase-error saturation phenomenology observed in precision oscillator characterization~\cite{rutman1978}.

For both free evolution and driven operations, the deleterious impact of using a  LO increases as the duration of the operation grows,
with relevance for all major qubit modalities, including trapped-ion~\cite{Monroe_Review}, superconducting~\cite{Oliver_MRS}, and semiconductor~\cite{Reilly_Rev} qubits.
%
%
For instance, for driven operations as long as $100\;\mu$s, particularly relevant for atomic qubits, the precision synthesizer only induces an error $\sim10^{-7}$ while the lab-grade synthesizer induces an error more than $20,000\times$ larger, reaching $\sim0.2\%$.
The primary reason for this behavior is the superior far-from-carrier phase-noise performance in the precision synthesizer, especially over the offfset frequency range $1-1000$ kHz.
We note that the selected extrapolation of the high-frequency phase noise beyond $\omega/2\pi=10$ MHz (see caption Fig.~\ref{Fig:InfidelityCurves}) is particularly favorable to the lab-grade LO, meaning that calculated error rates at the shortest times - $\tau < 100$ ns -- may underestimate the actual infidelities.

An interesting observation is that in the presence of these noise power spectral densities and for the realistic evolution times selected for our calculations, DES protocols have a minimal impact on gate performance. In Fig.~\ref{Fig:InfidelityCurves}c-d , over the entire range of times  -- 10 ns $\ldots$ 100 ms -- the use of these protocols offers only a small, sporadic improvement and, over several time spans, can even degrade the operational fidelity (note: these protocols may still give a substantial net improvement in the presence of more dominant environmental noise).  This performance is explained by considering that, while $\tilde{\mathcal{L}}(\omega)$ declines rapidly with offset frequency, the $\omega^2$ factor that transforms this spectrum to $S_{z}(\omega)$ reveals a high-frequency dominance of the resulting dephasing noise, in particular, for the precision LO, which exhibits an approximately Ohmic spectrum ($S_{z}(\omega) \propto \omega$, see Fig.~\ref{Fig:InfidelityCurves}b) ~\cite{Palma, Hodgson2010}.
DES is known to perform poorly for spectra with strong high-frequency content, where the noise evolves rapidly compared to the control and the physics of coherent averaging fails: a violation of the so-called decoupling limit.

\begin{table}[b]
\centering{}
\renewcommand{\arraystretch}{1.25}
    \begin{tabular}{|c|c|c|c|}
    %
    %
    \hline
    {\bf LO Class} & {\bf Operation} &{\bf Superconducting} & {\bf Trapped Ion}\\ \hline
    \multirow{2}{*}{Lab-Grade}
      & Primitive $\pi_{x}$ & $1.0\times10^{-6}$ & $2.7\times10^{-3}$ \\ 
      & WAMF $\pi_{x}$ & $4.4\times10^{-7}$ & $1.6\times10^{-3}$  \\ \hline
    \multirow{2}{*}{Precision} & Primitive $\pi_{x}$ & $1.6\times10^{-8}$ & $1.3\times10^{-7}$ \\ 
      & WAMF $\pi_{x}$ & $5.6\times10^{-8}$ & $4.9\times10^{-7}$  \\ \hline\hline
    {\bf Temperature} & {\bf Bandwidth} & \multicolumn{2}{c|}{{\bf Error Floor}}\\ \hline

    \multirow{3}{*}{290K (4K)} & 100 MHz & \multicolumn{2}{c|}{$6\times10^{-11}$($1\times10^{-12}$)} \\ 
      & 1 GHz  & \multicolumn{2}{c|}{$6\times10^{-10}$ ($1\times10^{-11}$)} \\ 
      & 10 GHz & \multicolumn{2}{c|}{$6\times10^{-9}$  ($1\times10^{-10}$)} \\ \hline
    \end{tabular}
    \caption{Comparison of typical error rates for superconducting (30~ns) vs trapped-ion (10~$\mu$s) driven-gate operations, including Walsh Amplitude Modulated Filters (WAMF).  Main results calculated assuming an integration bandwidth of 10 MHz.  The temperature-dependent noise-floor for different cutoff frequencies is also shown - these data are largely independent of operation time and type.  Noise floor calculations strictly valid only for cutoff frequencies high compared with the inverse operation time.}
\label{Table:T1}
\end{table}

\begin{figure}[t]
\centering
\includegraphics[width=0.9\columnwidth]{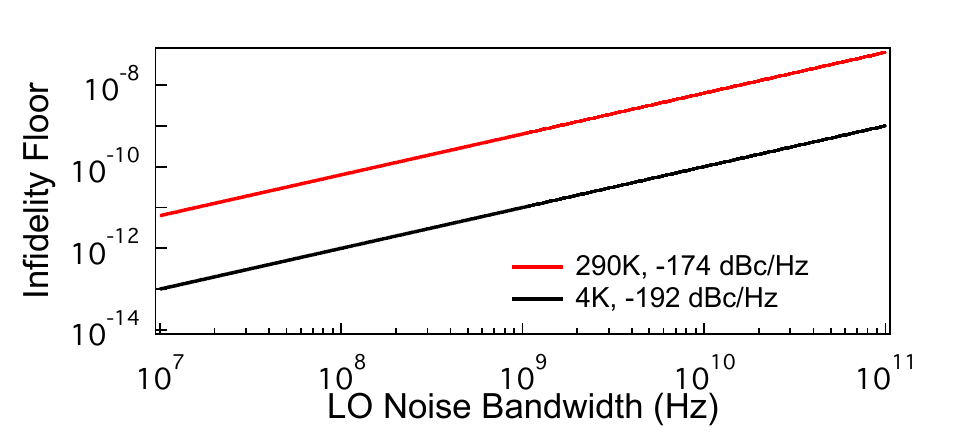}
\caption{Calculated infidelity floor (fidelity ceiling) imposed by thermal noise at 4K and 290K with variable LO noise bandwidths.  This calculation is independent of operation length, $\tau$ under the conditions outlined in the main text.  Results hold for Ramsey and both primitive and WAMF $\pi$ rotations.  Infidelity floor for spin echo is three times larger. See \emph{Supplementary Information}.}
\label{Fig:NoiseFloor}
\end{figure}

These results make clear that reducing far-from-carrier phase noise has the potential to provide augmented fidelities for quantum operations.  However it remains a question how well one might do by improving the quality of the LO.  We consider achievable gains in operational fidelity by calculating a lower-bound to the error imparted by the LO's thermal noise floor; even if LO hardware were improved we could do no better than saturating the thermal noise floor across the control bandwidth.  This is generically quoted as -174 dBm/Hz for a matched load at 290K.
This is an absolute noise power, meaning that for LOs with power 0 dBm the single-sideband phase noise floor will take a value $\tilde{\mathcal{L}}_{min} =-174$ dBc/Hz, and it will worsen (improve) with decreasing (increasing) LO power.
We assume an otherwise ideal LO subject to this minimum noise floor~\cite{ThermalNoise} by setting the phase noise to the constant value $\tilde{\mathcal{L}}(\omega) = \tilde{\mathcal{L}}_{min}$ over the effective control bandwidth $\omega\in[0,\omega_{c}]$. Using this spectrum we calculate the resulting upper-bound on operational fidelity in the presence of this noise using $\chi_{min}\approx(\kappa\omega_c/2\pi)10^{\tilde{\mathcal{L}}_{min}/10}$ (see \emph{Supplementary Information}).
In this expression $\kappa$ is a characteristic scaling factor depending on the control protocol, and $\chi_{min}$ is approximately independent of $\tau$ in the limit $\omega_{c}/2\pi\gg\tau^{-1}$.

This straightforward calculation reveals that broadband LO phase noise due to thermal effects at room temperature imposes a non-negligible upper bound on gate fidelity, as shown in Fig.~\ref{Fig:NoiseFloor}.
For instance, considering a typical bandwidth $\sim20$ GHz the thermal noise floor induces qubit infidelity in excess of $10^{-8}$.
Reducing the thermal noise floor to that associated with a 4K bath (-192 dBc/Hz) improves the fidelity by two orders of magnitude.  Similarly, restricting the control bandwidth to 10 MHz sets the infidelity $<10^{-11}$.

\section{Discussion}

The calculations we have presented demonstrate that master clock phase fluctuations are an emerging consideration for quantum information applications. Indeed, lab-grade oscillators may already limit the performance of today's qubits with gate times on microsecond scales.  On the other hand, using high-performance precision LOs, the calculated error rates for $\hat{\mathbb{I}}$ and $\hat{X}$ operations with both the fast and slow gates considered here are at several orders of magnitude smaller than the current state-of-art (Table 1). Therefore, although significantly more expensive, the use of such precision LOs is adequate for near-term, proof-of-concept experimental demonstrations with contemporary qubits.

Our results also enhance existing arguments about the merits of qubit modalities that accommodate fast control pulses. These arguments are typically based on a practical ``clock-speed'' assessment, wherein a technology with faster gates will simply execute an algorithm more quickly.  Here, in considering LO phase noise, we find that reducing operation times from $10\;\mu$s to 10 ns generally also reduces infidelity (Fig.~2 c-d). The improvement can be substantial (3-4 orders of magnitude) for lab-grade LOs, and it remains about an order of magnitude for the precision LO considered here. The intuition is that shorter pulses are subject to phase fluctuations for less time and, therefore, suffer less dephasing-induced errors.

However, arguments in favor of shorter control operations must be made with a cognizance of the actual LO noise spectrum and the control bandwidth. First, shorter operations access higher frequencies in the LO spectrum, and, as illustrated in Fig.~2, far-from-carrier phase noise dominates residual errors for commercial sources. Therefore, there may in principle be cases where decreasing the pulse duration actually increases the integrated noise level.  Second, bandwidth-dependent thermal-noise floors will eventually pose upper bounds on operational fidelities that are more strict when using short control pulses possessing higher bandwidths.  And, when they do, suppressing the effective thermal noise floor
by further engineering the performance of room-temperature LOs and/or developing cryo-compatible LOs~\cite{hartnett2010} embedded in cryogenic control architectures~\cite{hornibrook2015} may play an important role in the development of future, high-fidelity quantum information systems.

Similarly, whereas DES protocols are reasonably good at suppressing dephasing error due to low-frequency environmental fluctuations, they are less efficient at suppressing the impact of LO phase noise (Fig.~2a) due to the high-frequency weight in the dephasing power spectrum (Fig.~2b). In fact, over certain frequency bands in Figs.~2c-d, the DES protocols (dashed lines) can actually enhance error due to LO phase noise. Noise at higher frequencies evolves too rapidly to be coherently averaged by the control, resulting in the filter transfer functions for the DES protocols passing noise in this band.  The detailed spectrum ultimately determines whether shorter pulses and DES protocols yield a net win despite the higher noise far from the carrier.


Moving to mesoscale quantum information, such as quantum simulations and/or prototype logical-qubit demonstrations with a few hundred qubits~\cite{Wecker:2014bm,Bauer:2015tu,Reiher:2016ts}, we believe these results will serve to inform error budgets and motivate careful selection of the precision LO in use.  Looking forward, however, the bounds on error rates calculated here for both free and driven evolution with precision oscillators are, in the view of these authors, remarkably high and will likely limit performance in larger-scale systems, particularly in the context of quantum error correction.

%
It is widely viewed that quantum error correction, QEC, will be necessary to achieve sustained operation in faulty qubit systems.  Error bounds due to LO phase instability (of the type calculated here) place architectural constraints on the error correction protocol:  the QEC cycle must be completed within a time corresponding to a fixed phase-noise-induced error rate.  The tabulated maximum QEC cycle times appearing in Table~\ref{Table:T2}  show differences of more than $10^{5}$ in the permissible QEC cycle times for the two oscillators considered here.

The use of a precision LO in this context readily supports phase noise infidelities at the level $10^{-7}$, an order of magnitude below the most pessimistic fault-tolerance thresholds, over a time period $600\;\mu$s.
This provides strong evidence that, in principle, we should be able to achieve fault-tolerance for QEC with existing oscillator technology.
However, scalability requires consideration of corrected fault-tolerant error rates that are sufficiently low to permit algorithmic execution with manageable resource levels~\cite{NC}, a target that grows increasingly challenging for larger-scale applications~\cite{oskin2002practical}.
Broadly, one would aim to have a logical error rate that scales as the inverse problem size, with only a small chance of a single logical error during the algorithmic execution.  This results in calculated logical qubit error rates reaching $\mathcal{O}(10^{-15})$ in the context of Shor's algorithm for factorization of medium-sized keys~\cite{Kubiatowicz}.  

Achieving such extraordinary logical-error-rate targets for large-scale machines will involve a consideration of both the physical-qubit error rate and the resource levels required to implement QEC.
As error rates approach the fault-tolerance threshold, the encoding resource requirements diverge in time and qubit numbers.
Stated in an alternate way, the efficiency of QEC grows as the error rate is suppressed relative to the fault-tolerance threshold.  Accordingly, the error floors imposed by LO phase instability must be traded off at a system-level against the overheads associated with QEC encoding, for instance in the number of required physical qubits encoding a single logical bit.
Calculations suggest that for the Bacon-Shor code, the presence of clock-induced errors near $\sim10^{-7}$ would require a factor  $100-1000$ times more resources relative to physical-qubit error-rates near $10^{-10}$. While the LO is only part of the story -- errors due to environmental qubit dephasing must also be reduced -- one will also need to improve LOs to achieve these levels~\cite{Kubiatowicz}.
Based on current understanding of the relative challenge associated with QEC encoding in either large surface codes or concatenated schemes, and the continuing, anticipated improvements in physical qubit error rates, we believe that the clock-induced error rates we have identified here motivate investment in ultra-low-phase noise LO development.

\begin{table}[b]
\centering{}
\renewcommand{\arraystretch}{1.5}
    \begin{tabular}{|c|c|c|c|c|c|}
    \hline
    \multicolumn{6}{|c|}{{\bf Time to Reach LO-Induced Error Rate $p$ }} \\ \hline \hline
    $p$          & $10^{-3}$  & $10^{-4}$ & $10^{-5}$ & $10^{-6}$ & $10^{-7}$ \\ \hline
    Lab-Grade LO & 4.0 $\mu$s & 900 ns    & 200 ns    & 30 ns     & <10 ns     \\ 
    Precision LO & $>100$ ms  & $>100$ ms & $>100$ ms & 80 ms     & 600 $\mu$s \\ \hline
    \end{tabular}
\caption{Time until a qubit physical error rate $p$ is reached due solely to phase fluctuations in the LO. These times may be viewed as an upper-bound on the allowable QEC cycle period. Achievable cycle periods will be reduced due to other error sources. Error rates are derived from free-evolution calculations presented in Fig.~\ref{Fig:InfidelityCurves}b. Driven-operation error rates (see Fig.~\ref{Fig:InfidelityCurves}c) yield similar results.}
\label{Table:T2}
\end{table}

Another major consideration in the context of QEC is that clock-induced  errors are highly correlated in space and time, posing a challenge to existing QEC analyses which have focused primarily on independent and identically distributed stochastic error models.
As experimental teams incorporate clock distribution in quantum information architectures~\cite{gambetta2014frequency, hornibrook2015, DiCarlo2015}, the complexity of QEC analyses will need to be augmented in order to handle the effects of such correlated errors.

While we have focused on phase fluctuations far from the carrier -- i.e., on short time scales such as the physical qubit gate time or the QEC cycle period -- attention to long-term LO stability is also required.  Slow phase diffusion of the LO -- \emph{i.e.}, close to carrier phase noise -- causes substantial error accumulation over long times.
In the context of a single LO, such errors appear adiabatic and are generally correctable provided the timescale of the phase diffusion is much longer than the QEC cycle period.
By contrast, in systems with multiple control generators, long-time instabilities may be detrimental if the generators exhibit phase diffusion with respect to one-another.  In such cases, the diffusion represents a temporal, stochastic analog to the spatial, deterministic ``clock skew'' observed in classical semiconductor chips, and it is not generally accommodated by standard quantum error correction protocols.
To give a concrete example, whereas generators A and B may individually exhibit correctable slow phase diffusion, the axis for an $\hat{X}$ gate on generator A may adiabatically skew to be (in an extreme case) in quadrature with the equivalent $\hat{X}$ gate on generator B.  A logical controller directs generator A to apply an $\hat{X}$ gate to a qubit A, but the rotation is a $\hat{Y}$ rotation relative to generator B and the qubits it drives.

Temporally correlated, low-frequency LO phase fluctuations can also introduce challenges in evaluating relevant gate errors for QEC, due to the presence of potential biases in common quantum verification processes such as randomized benchmarking~\cite{HarrisonRB}.
In evaluating the impact of such errors, it is important to distinguish the contribution of slow LO instabilities to single-operation error rates from their contribution to empirical estimates of qubit fidelities drawn from tomographic measurements over repeated trials.
 The process of data acquisition and averaging over many individual experiments can contribute error due to slow drifts that dominate the actual error rate experienced in any individual operation or evolution period (see~\cite{Fei_2012} for details).

In summary, we have presented a comprehensive review of critical issues relating to master-clock-induced errors for quantum information applications.
We have unified established concepts from frequency metrology and quantum information, permitting experimentalists to translate common LO hardware specifications into estimates of qubit coherence and operational fidelities.
Analyses employing information available on representative LO datasheets have revealed that far-from-carrier phase noise poses a performance limiting upper-bound on operational fidelities.
%
As a result we expect to see a growing emphasis on high-performance LO synthesis chains across all technology platforms, and we foresee a growing importance of LO-induced errors in the design of large-scale quantum information systems.
We encourage future studies to examine not only clock synthesis, but clock distribution, with an eye towards architectural impacts of skewed clock distribution in quantum systems with single and multiple LOs.

\small
\section{Methods}
In the appropriate interaction picture, environmental dephasing processes and those induced by LO phase fluctuations are formally equivalent.
We begin with the physical system Hamiltonian ($\hbar=1$) in the laboratory frame for a qubit possessing a nominal transition (angular) frequency $\Qsplitting$, driven by a local oscillator with carrier frequency $\wLO$,
\begin{align}\label{Eq:HamiltonianLabFrame}
    H_S = &\frac{1}{2}\Qsplitting\sigz +\frac{1}{2}\delta\Qsplitting(t)\sigz\nonumber\\
    &+ \Omega(t)\cos\Big(\wLO t +[\phi_{C}(t)+\phi_{N}(t)]\Big)\sigx.
\end{align}
Appearing in this expression are the operators $\hat{\sigma}_{i}$ representing the Pauli matrices in the Cartesian coordinate basis.  The first term corresponds to the Hamiltonian of the qubit under free evolution, while the second term captures environmental noise (e.g. magnetic field fluctuations), which effectively change the nominal qubit transition frequency by an amount $\delta \omega(t)$ as
\begin{align}
\Qsplitting \to \Qsplitting+\delta\Qsplitting(t).
\end{align}
The third term corresponds to the qubit-field interaction with amplitude $\Omega(t)$ driven by a local oscillator with nominal carrier frequency $\wLO$.
The time-dependent phase of the LO is partitioned into a "control" component (desired phase evolution), $\phi_C(t)$,  and a ``noise'' component (unwanted phase fluctuations), $\phi_N(t)$. 

Moving to the interaction picture co-rotating with the carrier frequency and making the rotating-wave approximation we obtain
\begin{align}\label{CarrierPictureHamiltonianA}
H_I^{(\wLO)}&=\frac{1}{2}(\Qsplitting-\wLO)\sigz +\frac{1}{2}\delta\Qsplitting(t)\sigz\nonumber\\& +
\frac{1}{4}\Omega(t)
\Big\{
e^{-i[\phi_{C}(t)+\phi_{N}(t)]}\sigp+
e^{i[\phi_{C}(t)+\phi_{N}(t)]}\sigm
\Big\}
\end{align}
where we define the qubit ladder operators $\hat{\sigma}_{\pm} = \sigx\pm i\sigy$.

The LO phase fluctuation produces a time-dependent frequency detuning through its time derivative, $\fnoise(t)\equiv\dot{\phi}_N(t)$, effectively transforming the LO frequency as
\begin{align}
\wLO\to \wLO+\fnoise(t).
\end{align}
In order to make this phenomenology explicitly comparable with environmental dephasing, we perform a second interaction-picture transformation
\begin{align}
H_I^{(\wLO,\dot{\phi}_N)} \equiv U_{\dot{\phi}_N}^\dagger H_I^{(\wLO)} U_{\dot{\phi}_N} - \HphiN
\end{align}
where $U_{\dot{\phi}_N}(t) =\exp[-i\frac{\phi_N(t)}{2}\sigz]$ is the evolution operator under the dephasing Hamiltonian $\HphiN \equiv \frac{1}{2}\dot{\phi}_N(t)\sigz$ induced by phase fluctuations in the LO. Setting the static LO detuning to zero ($\Qsplitting-\wLO=0$) the transformed system Hamiltonian subject to LO phase fluctuations thereby takes the form
\begin{align}
H_I^{(\wLO,\dot{\phi}_N)}&= \frac{1}{2}\delta\Qsplitting(t)\sigz -\frac{1}{2}\dot{\phi}_N(t)\sigz\nonumber\\
\label{Eq:InteractionHamiltonianFinal}&+\frac{1}{2}\Omega(t)
\Big\{
\cos[\phi_C(t)]\sigx+
\sin[\phi_C(t)]\sigy
\Big\}.
\end{align}

\noindent The first two terms in this expression represent the two dephasing terms introduced in the main text, $\dh{z}^{(env)}(t) = \frac{1}{2}\delta\Qsplitting(t)$ and $\dh{z}^\text{(LO)}(t) =  -\frac{1}{2}\fnoise(t)=-\frac{1}{2}\dot{\phi}_N(t)$.

\acknowledgements
The authors acknowledge J. J. Bollinger for motivating careful consideration of the role of LO phase noise on qubit coherence and for discussions on phase noise.  We acknowledge useful conversations with S. Shankar and R. J. Schoelkopf, who provided motivation for the choice of lab-grade synthesizer.  Work partially supported by the ARC Centre of Excellence for Engineered Quantum Systems CE110001013, the Intelligence Advanced Research Projects Activity (IARPA) through the ARO, the US Army Research Office under Contract W911NF-12-R-0012, and a private grant from H. \& A. Harley.

\bibliography{ClockMainBIB}

\begin{thebibliography}{10}
\bibitem{Hamiltonian}
\bibinfo{note}{This Hamiltonian considers an ensemble of identically prepared
  noninteracting qubits subjected to a noisy environment and an external
  controller. Because we assume an interaction picture co-rotating with a
  resonant driving field, the freely evolving qubit Hamiltonian $\Hq \equiv
  \frac{1}{2}\omega_0\sigz$ does not appear explicitly (see Eq.
  \ref{Eq:HamiltonianLabFrame}).}
\bibitem{ThermalNoise}
\bibinfo{note}{Higher-power LOs can in principle have lower phase-noise floors,
  but will generally suffer from additional amplifier-induced phase noise. We
  therefore consider the thermal noise floor for a LO at 0 dBm output as an
  approximate lower-bound for the phase noise. We assume that the broadband
  thermal noise is actually manifested as temperature-dependent phase
  fluctuations in the LO rather than a background observed in the measurement
  process.}
\end{thebibliography}

\end{document}